\documentclass[conference]{IEEEtran}
\usepackage{array}
\usepackage{cite}
\usepackage{amsmath,amssymb,amsfonts}
\usepackage{algorithmic}
\usepackage{graphicx}
\usepackage{textcomp}
\usepackage{comment}
\usepackage{xcolor}
\usepackage{siunitx}
\usepackage{listings}
\usepackage{array}
\usepackage[most]{tcolorbox}
\tcbuselibrary{listings}
\usepackage[most]{tcolorbox}
\usepackage{graphicx}
\tcbuselibrary{listings}
\def\BibTeX{{\rm B\kern-.05em{\sc i\kern-.025em b}\kern-.08em
    T\kern-.1667em\lower.7ex\hbox{E}\kern-.125emX}}
\begin{document}

\title{Bridging the Gap: Empowering Small Models in Reliable OpenACC-based Parallelization via GEPA-Optimized Prompting}

\author{
\IEEEauthorblockN{Samyak Jhaveri}
\IEEEauthorblockA{School of Information and Computer Science \\
University of California, Irvine \\
\texttt{samyaknj@uci.edu}}
\and
\IEEEauthorblockN{Cristina V. Lopes}
\IEEEauthorblockA{School of Information and Computer Science \\
University of California, Irvine \\
\texttt{lopes@uci.edu}}
}
\maketitle

\begin{abstract}
Directive-based parallel programming frameworks like OpenACC lower the barrier to GPU-offloading by abstracting low-level programming details. However, manually writing high-performance pragma remains a significant challenge, which requires expertise in memory hierarchies, data movement, and parallelization strategies. Large Language Models (LLMs) present a promising potential solution for automated parallel code generation, but naive prompting often results in syntactically incorrect directives, uncompilable code, or performance that fails to exceed CPU baselines.

In this work, we present a systematic prompt optimization approach to enhance OpenACC pragma generation without the prohibitive computational costs associated with LLM post-training. We leverage the GEPA (GEnetic-PAreto) framework to iteratively evolve prompts through a reflective feedback loop. This process uses crossover and mutation of prompt instructions guided by expertly curating ``gold'' pragma examples and structured feedback based on clause and parameter-level mismatches between ``gold'' pragma and predicted pragma.

In our evaluation on the PolyBench suite, we observe a significant increase in compilation success rates for programs annotated with OpenACC pragma generated using the optimized prompts compared to those annotated using the simpler initial prompts, particularly for the smaller and cheaper "nano"-scale models. Specifically, with optimized prompts, the compilation success rate for GPT-4.1 Nano improved from $66.7\%$ to $93.3\%$, and for GPT-5 Nano it improved from $86.7\%$ to $100\%$, matching or surpassing the capabilities of their significantly larger and more expensive versions. Beyond compilation, the optimized prompts resulted in a $21\%$ increase in the number of programs that achieve functional GPU speedups over CPU baselines (i.e. $\text{speedup} > 1.0\times$). These results demonstrate that prompt optimization effectively unlocks the potential of smaller, cheaper LLMs to write stable and effective GPU-offloading directives, establishing a cost-effective pathway and lowering the expertise barrier to automated directive-based parallelization in HPC development workflows.

\end{abstract}



There is an ever-growing demand to maximize the utilization of the increasingly powerful and complex GPU hardware for High-Performance scientific, research, and commercial workloads. However, achieving strong parallelization speedups remains a significant bottleneck due to the specialized expertise required to manage memory hierarchies, data movement, and architecture-specific parallelization strategies. Directive-based frameworks such as OpenACC~\cite{openacc_specification_2021, parallel_programming_with_openacc_book} offer a practical middle-ground between performance and portability. By annotating their code with OpenACC's compiler directives (e.g., \texttt{\#pragma acc}), developers can iteratively offload compute--intensive regions to the GPU without rewriting entire applications. However, obtaining significant speedups still requires a nuanced understanding of the data dependencies to write the most effective clauses. This remains an error-prone, time-consuming and iterative process that relies on frequent profiling to refine data movement and parallelization strategies~\cite{automatic_parallelism_data_dependency, parallel_programming_with_openacc_book, mahmud_autoparllm_2025,unveiling_parallelism_in_serial_code,Li2013DiscoveryOP}. It becomes more difficult with increase in scale and complexity of the codebase often seen in many scientific HPC applications. Traditional automated tools, such as DawnCC~\cite{dawncc_2017} and KernelGen~\cite{Mikushin2014KernelGenT}, rely on brittle static analysis. They are also difficult to maintain, as they are built on top of fragile compiler infrastructure or require low-level compiler modifications.

\begin{figure*}
    \centering
    \includegraphics[width=\textwidth]{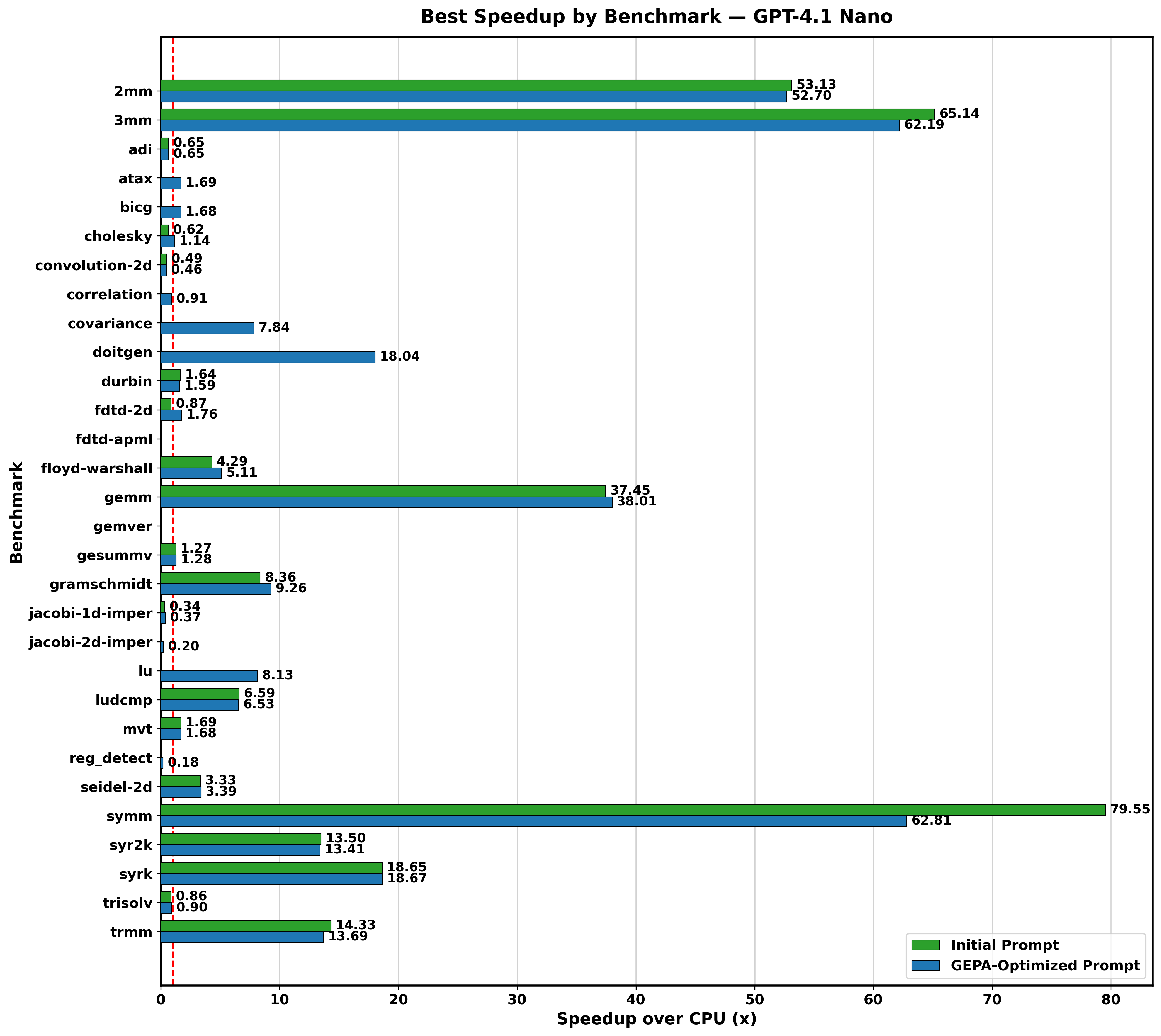}
    \caption{Speedup GPT-4.1 Nano}
    \label{fig:speedup_wall_best_gpt_4_1_nano}
\end{figure*}

\begin{figure*}
    \centering
    \includegraphics[width=\textwidth]{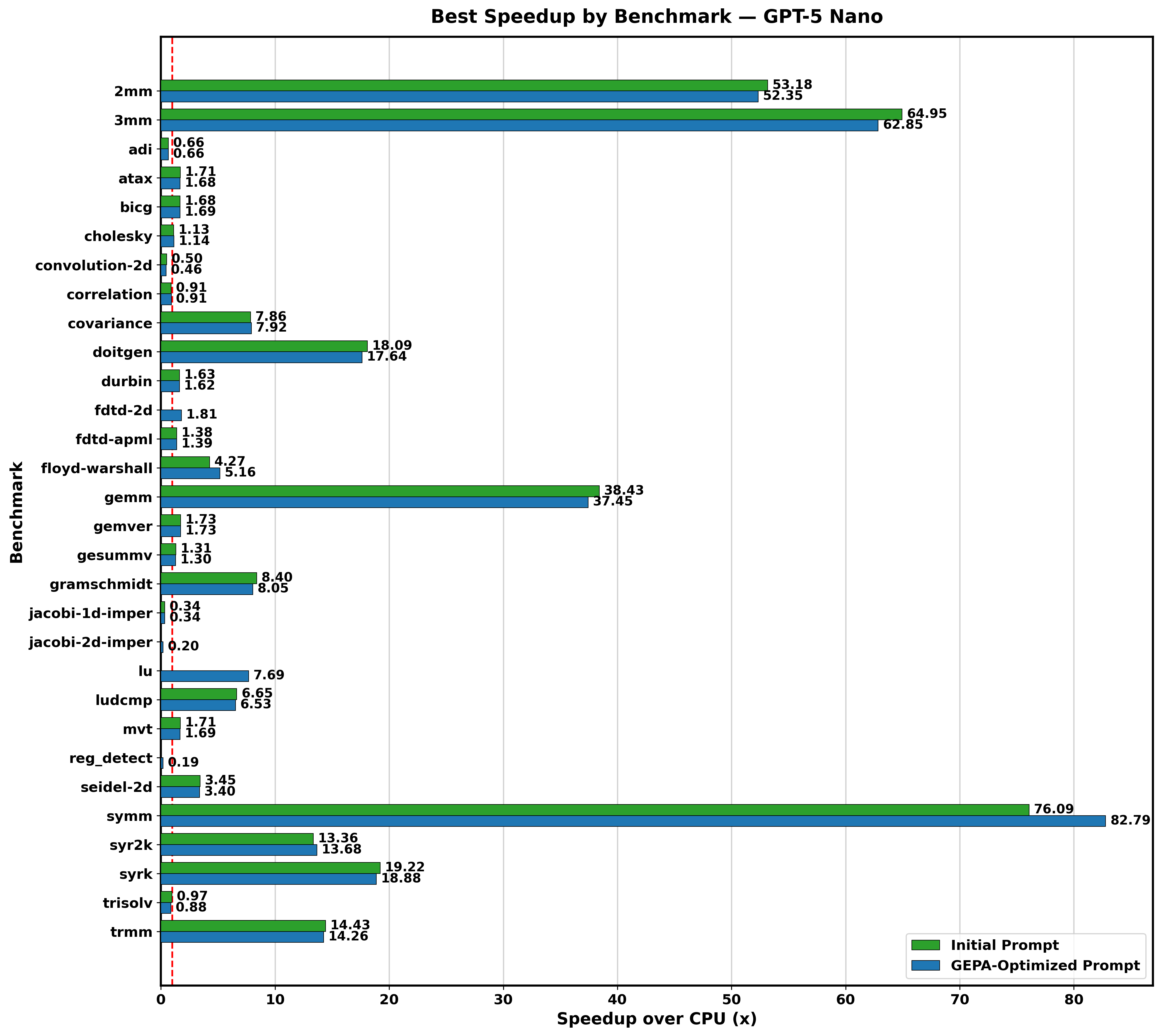}
    \caption{Speedup GPT-5 Nano}
    \label{fig:speedup_wall_best_gpt_5_nano}
\end{figure*}

Large Language Models (LLMs) have recently emerged as a promising solution for automated parallel code generation. We have observed that while large frontier models, like GPT-5 for example, demonstrate strong reasoning capabilities with a deep knowledge of OpenACC, still fail to produce compilable pragmas relevant to the entire context of a program. Furthermore, their high operational costs and latency often preclude their use in large-scale HPC development workflows. Conversely, smaller ``nano''--scale models, like GPT-4.1 Nano or GPT-5 Nano, are far more economically viable but fail to grasp the subtle nuances of data dependencies and requirements of parallel operations using OpenACC pragma, leading to poor compilability and suboptimal GPU speedups (Figs.~\ref{fig:speedup_wall_best_gpt_4_1_nano} and~\ref{fig:speedup_wall_best_gpt_5_nano}). Developers wanting to use LLMs for help with OpenACC are then faced with the dilemma of either using expensive new models that perform reasonably well, but not always correct, or using cheaper models that often generate the wrong pragmas.

Our work provides both a study of OpenAI's models' capabilities for OpenACC, as well as an approach that substantially improves these models' capabilities for generating correct and optimized OpenACC pragmas. In this work, we demonstrate that less powerful, cost-effective LLMs can achieve highly effective in-context knowledge of generating OpenACC pragma through systematic prompt optimization. We leverage the Genetic-Pareto (GEPA) framework~\cite{agrawal2025gepareflectivepromptevolution}, a reflective prompt optimizer that employs natural language reflection to learn high-level parallelization rules from trial and error. By evaluating the semantic correctness of the predicted pragma compared to the ``gold'' pragma, and issuing structured feedback at each iteration of the feedback loop, GEPA evolves a simple initial prompt into a highly optimized final prompts that specializes the ``nano'' models for OpenACC tasks.

Our evaluation using the PolyBench suite reveals a significant "robustness jump," with compilation success rate across all tested models increasing from 78.3\% to 95.8\%. The impact is most pronounced for small models: GPT-4.1 Nano’s compilation success rate improved from 66.7\% to 93.3\%, while GPT-5 Nano achieved a perfect 100\% success rate. Furthermore, the optimized prompts expanded the total number of benchmarks achieving functional GPU speedup by 21\%. Our results also show that optimized prompts enhance the ``nano''--scale models to surpass the capabilities of their larger, more expensive counterparts in generating correct pragmas that deliver substantial GPU speedups, such as the $82.79x\times$ and $38.01\times$ gains observed in the \texttt{symm}(Fig.~\ref{fig:speedup_wall_best_gpt_5_nano}) and \texttt{gemm}(Fig.~\ref{fig:speedup_wall_best_gpt_4_1_nano}) kernels, respectively. This work provides a scalable cost-effective path for automated HPC parallelization.

The primary contributions of this work are as follows:
\begin{itemize}
    \item We introduce a pragma normalization and comparison method that provides clause-level and parameter-level feedback to the LLM, along with structured feedback to guide the reflection model towards precise prompt mutations that correct complex data dependency and parallelization errors over the GEPA optimization iterations.
    \item We curated a dataset of 64 ``gold'' OpenACC examples, refined using NVIDIA Nsight Profiler, to serve as ground truth for training and evaluating the predictions made during the GEPA optimization process.
    \item We empirically demonstrate a "robustness jump" in compilation success rates, specifically for cost-effective ``nano''-scale models, with GPT-5 Nano achieving a 100\% success rate on the PolyBench suite. Furthermore, our optimized prompts enable smaller and cost-effective ``nano'' models to generate OpenACC pragma that produced speedups at par with those generated by their larger more expensive versions. 
\end{itemize}

The remainder of this paper is organized as follows: Section~\ref{sec:background_gepa} provides background on the GEPA prompt optimization framework. Section~\ref{sec:approach} details our Approach, including the integration of GEPA with OpenACC, the semantic similarity scoring metric, the two-stage inference process (Data Management and Loop Parallelization), and the curation of the ``gold'' pragma dataset. Section~\ref{sec:evaluation} describes the Evaluation methodology. Section~\ref{sec:results} presents the Results. Section~\ref{sec:related_work} discusses Related Work in automatic pragma generation and LLM-driven synthesis. Section~\ref{sec:limitations} addresses the Limitations of the current study. Section~\ref{sec:conclusion_and_future_work} concludes the paper and outlines the directions for Future Work.

\section{Background: GEPA Prompt Optimization}
\label{sec:background_gepa}
GEPA (\textbf{Ge}netic-\textbf{Pa}reto)~\cite{agrawal2025gepareflectivepromptevolution} is a prompt optimization framework that employs a combination of natural-language feedback-driven reflection and multi-objective evolutionary search to improve the LLM performance. This approach treats prompts as evolvable elements and systematically refines them based on execution traces and evaluation metrics. These metrics can be customized to serve the needs of generation application. GEPA optimization relies on three main principles: reflective mutation, evolutionary candidate pool, and Pareto-optimal selection. A lower-parameter, cost-effective ``student" model serves as the inference engine. An LLM with strong reasoning capabilities (like GPT-5) is used as a ``reflection" model to analyze execution traces, including reasoning steps and metric evaluation results of the responses generated by the candidate ``student" model. The reflection model uses targeted, natural language feedback from humans, which is then used to generate a new, mutated prompt variant, thereby learning from trial-and-error.

The process of prompt optimization involves the management of a diverse population of prompt candidates, iteratively evolving them over multiple generations through mutation, and merging the successful candidates. In each iteration, GEPA identifies all the prompts that achieve the highest score on at least one task. The framework maintains a Pareto frontier of candidate prompts that perform best, to avoid local optima. This approach ensures that a diverse set of promising prompts is maintained. To evolve the current prompts into new prompts, GEPA stochastically explores prompts that excel across different problem instances. A prompt stays on the Pareto frontier if no other prompt beats it on every objective. If Candidate A is better than or equal to Candidate B across all instances and strictly better on at least one, Candidate B is "dominated" and removed from the pool to focus resources on more promising candidates. Once the optimization budget is exhausted, the final ``best'' candidate is typically chosen by ranking all prompts on the frontier by their aggregate (mean) score across the entire validation set. This ensures that the final prompt generalizes well by combining lessons from multiple successful ancestors rather than just refining a single initial approach.

This methodology contrasts with traditional Reinforcement Learning (RL) methods by leveraging the interpretable nature of language as a learning signal, often achieving superior performance with significantly fewer rollouts~\cite{agrawal2025gepareflectivepromptevolution}.

\section{Approach}
\label{sec:approach}

\subsection{GEPA for OpenACC Pragma Generation}
\label{subsec:gepa_prompt_optimization}

Our approach leverages the GEPA framework within the DSPy library to evolve simple initial prompts (see Appendix, List.~\ref{lst:initial_data_management_prompt} and List.~\ref{lst:initial_loop_parallelization_prompt}) to generate optimized prompts that enable the lower-parameter and smaller ``nano'' models to generate syntactically and semantically correct OpenACC pragma for a given C/C++ program.

In each iteration, the ``student'' model prompted with a simple initial prompt and an example program with a single \texttt{\textless DM\_PRAGMA\textgreater} or \texttt{\textless LP\_PRAGMA\textgreater} tag to indicate the what kind of pragma it needs to generate. The quality of the predicted pragma is determined by semantically comparing it with the ``gold'' pragma written by us for that particular tagged location in the program. The pragma are also normalized to a canonical map representation to compare them semantically and identify clause-level and parameter-level mismatch between the predicted pragma and the ``gold'' pragma. The core of our optimization strategy lies in the structured granular feedback report detailing clause- and parameter-specific mismatches. This feedback, along with the semantic similarity score, is passed on to the GEPA optimizer for the reflection model. The reflection model uses this feedback to mutate the prompt towards better performance in the next iteration, i.e., getting the model to predict pragma that are semantically more similar to the corresponding ``gold'' pragma for a particular tagged location in the program. 

\subsection{Semantic Similarity Scoring}

At the heart of GEPA there is the automatic comparison between predicted pragma and ``gold'' pragma (ground truth). Clearly, evaluating the accuracy of generated OpenACC pragmas via direct textual comparison is insufficient. Standard string-matching metrics such as BLEU\cite{bleu_score} or Exact Match are order-sensitive and do not account for the commutativity of clauses (e.g., $\texttt{copy(A) copyin(B, C)} \equiv \texttt{copyin(B, C) copy(A)}$) and the order-independence of variable lists within a clause. To address this, we define a normalization function $N : \Sigma^* \to \mathcal{M}$ that maps a raw pragma string to a structured canonical representation $\mathcal{M}$.

The pragma normalization process involves three stages. First, the primary directive (e.g., \texttt{parallel loop}, \texttt{kernels}, \texttt{data}, etc.) is extracted from the raw pragma string. Any mismatch between the predicted and ground truth directive directly results in a penalty. Second, the remaining pragma string is split into its constituent clauses using a robust splitting algorithm that respects nested parentheses to handle complex expressions (e.g., array slicing \texttt{A[0:N]}). Third, for each clause $c$, the argument list $P_{c} = {p_{1}, p_{2}, ..., p_{n}}$ is sorted lexicographically and the whitespace amongst the clause arguments is normalized. For the \texttt{reduction} clause, special handling is applied by parsing the operator and variable list into a set of tuples $\{(op, var_{1}), (op, var_{2}), ...\}$, ensuring that \texttt{reduction(+:a, b)} and \texttt{reduction(+:b,a)} map to the same canonical representation. For example, as shown in Fig.~\ref{fig:normalization_example}, the different pragma input strings map to the same canonical representation.
\[
\begin{aligned}
\mathcal{M} = \{ & \text{``parallel loop''}, \\
                 & \text{private} \to (\text{``i''}), \\
                 & \text{reduction} \to (\text{``+ : sum''},\ \text{``+ : temp''}) \}
\end{aligned}
\]

\begin{figure}[!t]
\centering
\fbox{
\begin{minipage}{0.95\columnwidth}
\small
\textbf{Input (Gold):} \\
\texttt{\#pragma acc parallel loop reduction(+:sum, temp) private(i)}

\vspace{0.5em}
\textbf{Input (Predicted):} \\
\texttt{\#pragma acc parallel loop private(i) reduction(+:temp, sum)}
\end{minipage}
}
\caption{Comparison of Gold and Predicted OpenACC Pragma.}
\label{fig:normalization_example}
\end{figure}

To measure the semantic similarity between predicted and ground truth pragma, the normalized predicted pragma map $P$ is compared against the gold standard pragma map $G$. Clause-level F1 and parameter-level F1 scores are calculated by treating the clause names (the keys of the maps) as sets $K_{G}$ and $K_{P}$. Let $I = K_{G} \cap K_{P}$, represent the intersection of the clause sets.
\[
\mathrm{Precision}_{\textit{clause}}=\frac{\lvert I\rvert}{\lvert K_P\rvert}, 
\qquad
\mathrm{Recall}_{\textit{clause}}=\frac{\lvert I\rvert}{\lvert K_G\rvert}
\]

\[
\mathrm{F1}_{\textit{clause}}
=2\cdot
\frac{\mathrm{Precision}_{\textit{clause}}\cdot \mathrm{Recall}_{\textit{clause}}}
{\mathrm{Precision}_{\textit{clause}}+\mathrm{Recall}_{\textit{clause}}}
\]

For the subset of clauses within the intersection $I$, we evaluate parameter accuracy. Let $v_{G,c}$ and $v_{P,c}$ be the multi-sets of normalized parameters for clause $c$ in the gold and predicted pragma maps, respectively. The total matches (Hits) are calculated across all shared clauses as: 
\[
\mathrm{Hits}=\sum_{c\in I}\left\lvert v_{G,c}\cap v_{P,c}\right\rvert
\]

\[
\mathrm{Precision}_{\textit{param}}
=\frac{\mathrm{Hits}}{\sum_{c\in I}\left\lvert v_{P,c}\right\rvert},
\qquad
\mathrm{Recall}_{\textit{param}}
=\frac{\mathrm{Hits}}{\sum_{c\in I}\left\lvert v_{G,c}\right\rvert}
\]

The final score, $\mathrm{S}_{\textit{total}}$, is a weighted sum that prioritizes structural correctness while heavily penalizing incorrect data movement or privatization variables.
\[
\mathrm{S}_{\textit{total}} = 0.6 \cdot \mathrm{F1}_{\textit{clause}} + 0.4 \cdot \mathrm{F1}_{\textit{param}}
\]

\noindent
\textbf{Scoring Examples:} 
The following cases illustrate the metric's behavior across different error types:

\textbf{Perfect Semantic Match (Fig.~\ref{fig:perfect_semantic_matching_example})}: Predicted and gold pragma differ only in clause order.
$$\mathrm{F1}_{\textit{clause}} = 1.0, \mathrm{F1}_{\textit{param}} = 1.0 \Rightarrow \mathrm{S}_{\textit{total}} = 1.0$$ 

\textbf{Correct Structure, Incorrect Variable (Fig.~\ref{fig:correct_structure_incorrect_variable_example}):} The clause sets match ($\mathrm{F1}_{\textit{clause}} = 1.0$), but the parameter $j$ is missing. $\mathrm{Precision}_{\textit{param}} = 1.0$, $\mathrm{Recall}_{\textit{param}} = 0.5 \Rightarrow \mathrm{F1}_{\textit{param}} \approx 0.66$. Total score: $$\mathrm{S}_{\textit{total}} = 0.6 \cdot (1.0) + 0.4 \cdot (0.66) = 0.86$$ 

\textbf{Structural Mismatch, Incorrect Primary Directive (Fig.~\ref{fig:incorrect_primary_directive_example})}: A mismatch in the primary directive triggers significant penalty due to structural divergence from gold standard, resulting in $\mathrm{F1}_{\textit{clause}} \approx 0.5$.

\begin{figure}[!t]
\centering
\fbox{
\begin{minipage}{0.95\columnwidth}
\small
\textbf{Gold:} \\
\texttt{\#pragma acc data copyin(A[0:N]) copyout(B[0:N])}

\vspace{0.5em}
\textbf{Predicted:} \\
\texttt{\#pragma acc data copyout(B[0:N]) copyin(A[0:N])}
\end{minipage}
}
\caption{Example of Perfect Semantic Matching.}
\label{fig:perfect_semantic_matching_example}
\end{figure}

\begin{figure}[!t]
\centering
\fbox{
\begin{minipage}{0.95\columnwidth}
\small
\textbf{Gold:} \\
\texttt{\#pragma acc parallel loop private(i, j)}

\vspace{0.5em}
\textbf{Predicted:} \\
\texttt{\#pragma acc parallel loop private(i)}
\end{minipage}
}
\caption{Example of Correct Structure, but Incorrect Variables.}
\label{fig:correct_structure_incorrect_variable_example}
\end{figure}

\begin{figure}[!t]
\centering
\fbox{
\begin{minipage}{0.95\columnwidth}
\small
\textbf{Gold:} \\
\texttt{\#pragma acc kernels copy(A)}

\vspace{0.5em}
\textbf{Predicted:} \\
\texttt{\#pragma acc parallel loop copy(A)}
\end{minipage}
}
\caption{Example of Incorrect Primary Directive.}
\label{fig:incorrect_primary_directive_example}
\end{figure}

\subsection{Inference}
Inference is conducted in two stages:
\begin{itemize}
    \item \textbf{Stage 1: Data Management (DM) Pragma Synthesis:} The model is prompted with a data management prompt to generate data management pragma for the \texttt{\textless DM\_PRAGMA\textgreater} sites in the serial PolyBench program. The synthesized DM pragmas from Stage 1 are re-inserted into the source, forming the input context for the second stage.
    
    \item \textbf{Stage 2: Loop Parallelization (LP) Pragma Synthesis:} The model is prompted with a loop parallelization prompt to generate compute pragma at the \texttt{\textless LP\_PRAGMA\textgreater} sites.
\end{itemize}

By performing loop-parallelization inference in the presence of the model-generated data management pragma rather than ``gold'' ground truth pragma, the evaluation accurately reflects real-world downstream usage. This approach ensures that the generated compute pragmas are contextually aware of the memory orchestration established in the preceding stage.

\begin{table*}
\centering
\caption{Feedback Analysis from GEPA Script}
\label{tab:gepa_feedback}
\scriptsize
\begin{tabular}{|p{3.5cm}|p{7cm}|p{4.5cm}|}
\hline
\textbf{Error Category} & \textbf{Prompt Hint} & \textbf{Corrective Action} \\ \hline
Missing collapse clause & The model failed to recognize that the nested loops are tightly coupled and can be collapsed. Ensure the prompt emphasizes checking for 'tightly nested loops' (loops with no intervening code) and apply 'collapse(N)' to maximize parallelism. & Add collapse(\{inner\}) clause. \\ \hline
Unnecessary collapse clause & The prompt must explicitly forbid 'collapse' if there are intervening statements or complex index dependencies between loops. & Remove collapse(\{inner\}) clause. \\ \hline
Missing reduction clause & The prompt should define a 'reduction' as a scalar accumulated across the loop (e.g. sum+=, max=) that is not re-initialized inside. Do not reduce arrays. & Add reduction(\{op\}:\{vars\_txt\}) clause. \\ \hline
Unnecessary, extra reduction clause & The prompt must clarify that reduction is ONLY for scalars that are truly accumulated across the loop, and not reinitialized per outer iterations, and not for arrays or private vars. Use the correct operator for the accumulated scalar. Do not reduce arrays. & Remove reduction(\{op\}:\{vars\_txt\}) clause. \\ \hline
Missing private clause & The prompt should explicitly state that scalars assigned within the loop body must be marked 'private' unless they are reductions to prevent data races. & Add private(\{inner\}) clause. \\ \hline
Extra or incorrect private clause & The prompt must state that loop iterators are implicitly private and read-only shared variables do not need privatization. & Remove private(\{inner\}) clause. \\ \hline
Missing present clause & The prompt must instruct the model to use 'present' for arrays that are already resident on the GPU (e.g., passed from a calling function or managed by an enclosing data region). & Add present(\{inner\}) clause. \\ \hline
Unnecessary present clause & The prompt should not use 'present' if the data is not guaranteed to be on the device. Use data movement clauses (copy/copyin) if transfer is needed. & Remove present(\{inner\}) clause. \\ \hline
Missing copyin clause & The prompt must enforce 'copyin' for arrays that are Read-Only inside the region and require initialization from the host. & Add copyin(\{inner\}) clause. \\ \hline
Unnecessary copyin clause & Do not use 'copyin' if the variable is written to (use copy/copyout) or not used at all. & Remove copyin(\{inner\}) clause. \\ \hline
Missing copyout clause & The prompt must enforce 'copyout' for arrays that are Write-Only on the device and whose results are needed back on the host. & Add copyout(\{inner\}) clause. \\ \hline
Unnecessary copyout clause & Do not use 'copyout' if the variable needs initial values from the host (use copy) or is not used. & Remove copyout(\{inner\}) clause. \\ \hline
Missing copy clause & The prompt must enforce 'copy' for arrays that are Read-Write (accessed and modified) inside the region. & Add copy(\{inner\}) clause. \\ \hline
Unnecessary copy clause & Use more specific clauses if possible: 'copyin' for Read-Only, 'copyout' for Write-Only. Use 'copy' only for true Read-Write dependencies. & Remove copy(\{inner\}) clause. \\ \hline
Missing create clause & The prompt must use 'create' for temporary arrays used only on the device (scratchpad) that do not require host values. & Add create(\{inner\}) clause. \\ \hline
Unnecessary create clause & Do not use 'create' if the variable needs initialization from the host (use copyin/copy). & Remove create(\{inner\}) clause. \\ \hline
collapse clause parameter mismatch & The prompt should instruct the model to list all relevant variables for the clause and verify against the variable declarations. & Use 'collapse(N)' as seen in {GOLD}). \\ \hline
reduction clause parameter mismatch & The prompt should instruct the model to list all relevant variables for the clause and verify against the variable declarations. Use reduction clause for the scalar that is truly accumulated across the parallelized dimension and not reinitialized per outer iteration. Do not reduce arrays. Do not duplicate variables in private() if they appear in reduction(). & Use reduction(\{gop\}:\{gvars\}) (as seen in GOLD), instead of reduction(\{pop\}:\{pvars\}) (currently in PRED) \\ \hline
private clause parameter mismatch & The prompt should instruct the model to list all relevant variables for the clause and verify against the variable declarations. Use reduction clause for the scalar that is truly accumulated across the parallelized dimension and not reinitialized per outer iteration. Do not reduce arrays. Do not duplicate variables in private() if they appear in reduction(). List only scalars written inside the loop and not covered by reduction(); loop indices are implicit private. & Use private(\{g\_inner\}) (as seen in GOLD), instead of private(\{p\_inner\}) (currently in PRED) \\ \hline
\end{tabular}
\end{table*}





\subsection{Curating High-Quality Examples}
\label{subsec:curating_high_quality_examples}

We constructed a dataset of C/C++ programs with high-quality OpenACC programs to use as ``gold'' examples for the GEPA prompt optimization process.

Potential candidates were identified via the GitHub Code Search API, targeting repositories with programs containing \texttt{\#pragma acc loop} and \texttt{\#pragma acc parallel loop} directives. Only programs that successfully compiled using the NVIDIA \texttt{nvc} or \texttt{nvcc} compilers were retained. We excluded compiler unit tests (e.g., GCC, LLVM) and standard benchmarks (e.g., PolyBench). To reduce noise, all comments and non-essential metadata were stripped from the source files. 

From the filtered corpus, we selected 64 programs based on computational density and potential to benefit most from GPU acceleration. We iteratively refined the pragmas for these programs using feedback from NVIDIA Nsight Profiler to ensure optimal data movement and loop scheduling. 

Once the pragma offered satisfactory speedups, these ``gold'' pragmas serve as the ground truth. 

The dataset we created has two sets of examples, one set focusing on data management pragma, and another one focusing on loop parallelization pragma. We parse the selected programs and categorize their existing pragma into one of the two categories. The data management pragma are identified by strings like \texttt{\#pragma acc data}, \texttt{enter data}, \texttt{update}, etc. The loop parallelization pragma are identified by strings like \texttt{\#pragma acc parallel loop}, \texttt{kernels}, etc. Each data management pragma entry in the dataset has the ``gold'' pragma, and the original source program with a \texttt{<DM\_PRAGMA>} tag replacing it in its original location in the program. All other pragma are removed. Each loop parallelization pragma entry has the ``gold'' pragma and the source program with \texttt{\textless LP\_PRAGMA\textgreater} tag replacing it in its original location in the program, with the ``gold'' data management pragma already present. Including the ``gold'' data management pragma in the source program provides context about the state of the data region already established in the program. This structured approach allows GEPA to learn the dependency between data movement and compute offloading, mirroring the manual optimization workflow of HPC developers.


\section{Evaluation}
\label{sec:evaluation}

We evaluated the compilation success rate and speedup over CPU baselines on the PolyBench suite of benchmarks~\cite{polybench_grauer2012auto} with OpenACC pragma generated by the models using the initial simple prompt as the baseline and the GEPA-optimized prompt (see Appendix~\ref{appendix_prompts} Figures~\ref{lst:initial_data_management_prompt} and ~\ref{lst:initial_loop_parallelization_prompt}). 
We selected the PolyBench/C 4.2.1~\cite{polybench_grauer2012auto} suite as our primary evaluation vehicle because it provides a standardized, diverse set of kernels representing fundamental computational patterns in scientific computing, such as stencil operations, linear algebra, and data mining. Although PolyBench consists of relatively small kernels, it serves as a rigorous baseline to isolate the ability of an LLM to correctly identify data dependencies and manage memory movement without the confounding complexity of large-scale inter-procedural analysis. By demonstrating success on these "building blocks" of HPC, we establish a performance ceiling for prompt-optimized directive generation before extending the methodology to more complex scientific mini-apps.\\
\textbf{Hardware Setup:} All GPU execution tests and performance measurements were conducted on a workstation equipped with an NVIDIA GeForce RTX 4070 GPU (12GB GDDR6X VRAM, Ada Lovelace architecture). The host system featured an AMD Ryzen 9 7900X 12-Core Processor with 32GB of system memory. We utilized the NVIDIA HPC SDK (nvc 24.5) for compilation, targeting the cc89 compute capability with the \texttt{-acc -fast} optimization flags. 



\section{Results}
\label{sec:results}

\subsection{Prompt Optimization Delivers a Robustness Jump}
GEPA-optimized prompts enhance the models' ability to generate syntactically and semantically correct OpenACC pragma, significantly improving the fraction of benchmarks that compiled successfully, relative to the fraction of benchmarks with pragma generated using the initial prompt with the same models. Across 120 model--benchmark evaluations (30 benchmarks from the PolyBench suite and four model variants: GPT-4.1, GPT-4.1 Nano, GPT-5, and GPT-5 Nano, best-of-5 runs per setting), aggregate compilability rose from $78.3\%$ ($94/120$) using the initial prompt to $95.8\%$ ($115/120$) with the GEPA-optimized prompt (Table~\ref{tab:compilability_speedup}). The use of the GEPA-optimized prompt converted 21 previously failing cases into successful GPU compilations and achieved zero regressions i.e. no benchmark that compiled under the initial prompt failed under the optimized prompt. 

As shown in Table~\ref{tab:per_model_compilability_speedup} the performance gains are consistent across all model sizes and are particularly pronounced for smaller, lower-capacity models. For ``nano''--scaled models, the optimized prompt significantly bridged the compilation reliability gap. The compilation success rate for GPT-4.1 Nano improved from $66.7\%$ ($20/30$) to $93.3\%$ ($28/30$), and for GPT-5 Nano from $86.7\%$ ($26/30$) to $100\%$ ($30/30$). A similar trend was observed for larger models, where failures were reduced to a smaller subset of highly complex benchmarks. The compilation success rates for GPT-4.1 improved from $83.3\%$ ($25/30$) to $96.7\%$ ($29/30$), and for GPT-5 from $76.7\%$ ($23/30$) to $93.3\%$ ($28/30$). Overall, using the optimized prompt reduces syntactic/semantic errors in OpenACC pragmas even when generated using the smaller model. This stabilization of offloading patterns is critical for automated HPC pipelines, where compilability serves as a primary metric for practical utility.

\begin{table}
    \centering
    \caption{Overall impact of optimized prompt on robustness in terms of Compilability and Speedup.}
    \label{tab:compilability_speedup}
    \resizebox{\columnwidth}{!}{%
    \begin{tabular}{|l|r|r|r|}
        \hline
        \textbf{Prompt} & \textbf{Compilable} & \textbf{Compilability Rate} & \textbf{Speedup$>1$ count} \\
        \hline
        Initial          & 94  & 78.3\% & 67 \\
        Optimized (ours) & \textbf{115} & \textbf{95.8\%} & \textbf{81} \\
        \hline
    \end{tabular}%
    }
    \\[2pt]
\end{table}

Despite the improvements made by using optimized prompts, a small subset of benchmarks remains intractable. As shown in Table~\ref{tab:rescued-failing} these failures are mainly concentrated in specific kernels. The \texttt{gemver} benchmarks failed to compile under both initial and optimized prompts for GPT-4.1 and GPT-4.1 Nano models. Furthermore, \texttt{fdtd-apml} was unsuccessful for GPT-4.1 Nano in both prompt configurations. For GPT-5 (Fig.~\ref{fig:speedup_wall_best_gpt_5}), the remaining failures are \texttt{floyd-warshall} and \texttt{jacobi-2d-imper}, which produce no valid GPU result under either prompt (Table~\ref{tab:rescued-failing}).

\begin{table}
    \centering
    \caption{Per-model compilability and speedup summary.}
    \label{tab:per_model_compilability_speedup}
    \setlength{\tabcolsep}{4pt}  
    \renewcommand{\arraystretch}{1.2}  
    \footnotesize  
    \resizebox{\linewidth}{!}{%
    \begin{tabular}{|l|l|c|c|c|}
        \hline
        \textbf{Model} & \textbf{Prompt} &
        \multicolumn{1}{c|}{\shortstack{\textbf{Compilability} \\ \textbf{(out of 30)}}} &
        \multicolumn{1}{c|}{\shortstack{\textbf{\# Benchmarks} \\[-0.2em] \textbf{with Speedup $>1$}}} \\
        \hline
        GPT-4.1      & Initial          & 83.3\% (25)  & 16 \\
        GPT-4.1      & Optimized        & 96.7\% (29)  & 22 \\
        GPT-4.1 Nano & Initial          & 66.7\% (20)  & 14 \\
        GPT-4.1 Nano & Optimized        & 93.3\% (28)  & 21 \\
        GPT-5 Nano   & Initial          & 86.7\% (26)  & 21 \\
        GPT-5 Nano   & Optimized        & 100.0\% (30) & 23 \\
        GPT-5        & Initial          & 76.7\% (23) & 16 \\
        GPT-5        & Optimized        & 93.3\% (28) & 15 \\
        \hline
    \end{tabular}
    }
\end{table}
\subsection{Additional Compilation Rate Means More Chances of Functional GPU Acceleration}
Among the 21 ``rescued'' cases (failed with the initial prompt, compiled successfully with the GEPA-optimized prompt), over $50\%$ ($11/21$) achieved functional speedup ($\text{speedup} > 1\times$). The total of accelerated model-benchmark pairs across the suite expanded by $21\%$ (from 67 to 81). High-impact rescues are presented in Table~\ref{tab:rescued-failing}. For example, GPT-4.1 Nano rescues \texttt{doitgen} to $18.03\times$,\texttt{covariance} to $7.83\times$, and \texttt{lu} to $8.12\times$ speedups which it failed when using the initial prompt. GPT-5 Nano rescues \texttt{fdtd-2d} to $1.81\times$ and \texttt{lu} to $7.68\times$, which it failed when using the initial prompt. GPT-4.1 rescues \texttt{gramschmidt} to $8.63\times$ and \texttt{fdtd-apml} to $1.35\times$. For GPT-5 (Fig.~\ref{fig:speedup_wall_best_gpt_5}), the optimized prompt enables \texttt{gramschmidt} at $4.9\times$ where the initial prompt produced no valid compilable OpenACC pragma. In practical terms, these examples show that improved compilability frequently translates into useful GPU performance rather than merely producing runnable-but-slower kernels.

\begin{table}
    \centering
    \caption{Speedup on the Common Compiled Subset of Benchmarks.}
    \label{tab:mean_speedup}
    \setlength{\tabcolsep}{4pt}  
    \renewcommand{\arraystretch}{1.2}  
    \footnotesize  
    \resizebox{\linewidth}{!}{%
    \begin{tabular}{|l|c|c|}
        \hline
        \textbf{Model} & 
        \multicolumn{1}{c|}{\shortstack{\textbf{Mean Speedup} \\ \textbf{(Initial Prompt})}} & 
        \multicolumn{1}{c|}{\shortstack{\textbf{Mean Speedup} \\ \textbf{(Optimized} Prompt)}} \\ 
        \hline
        GPT-4.1      & 2.401 & 3.669 \\
        GPT-4.1 Nano & 4.308 & 4.606 \\
        GPT-5 Nano   & 4.143 & 3.828 \\
        GPT-5        & 2.820 & 2.542 \\
        \hline
    \end{tabular}
    }
\end{table}

\subsection{Prompt Optimization Preserves Speedups While Improving Compilation Reliability}
\begin{figure*}
    \centering
    \includegraphics[width=\textwidth]{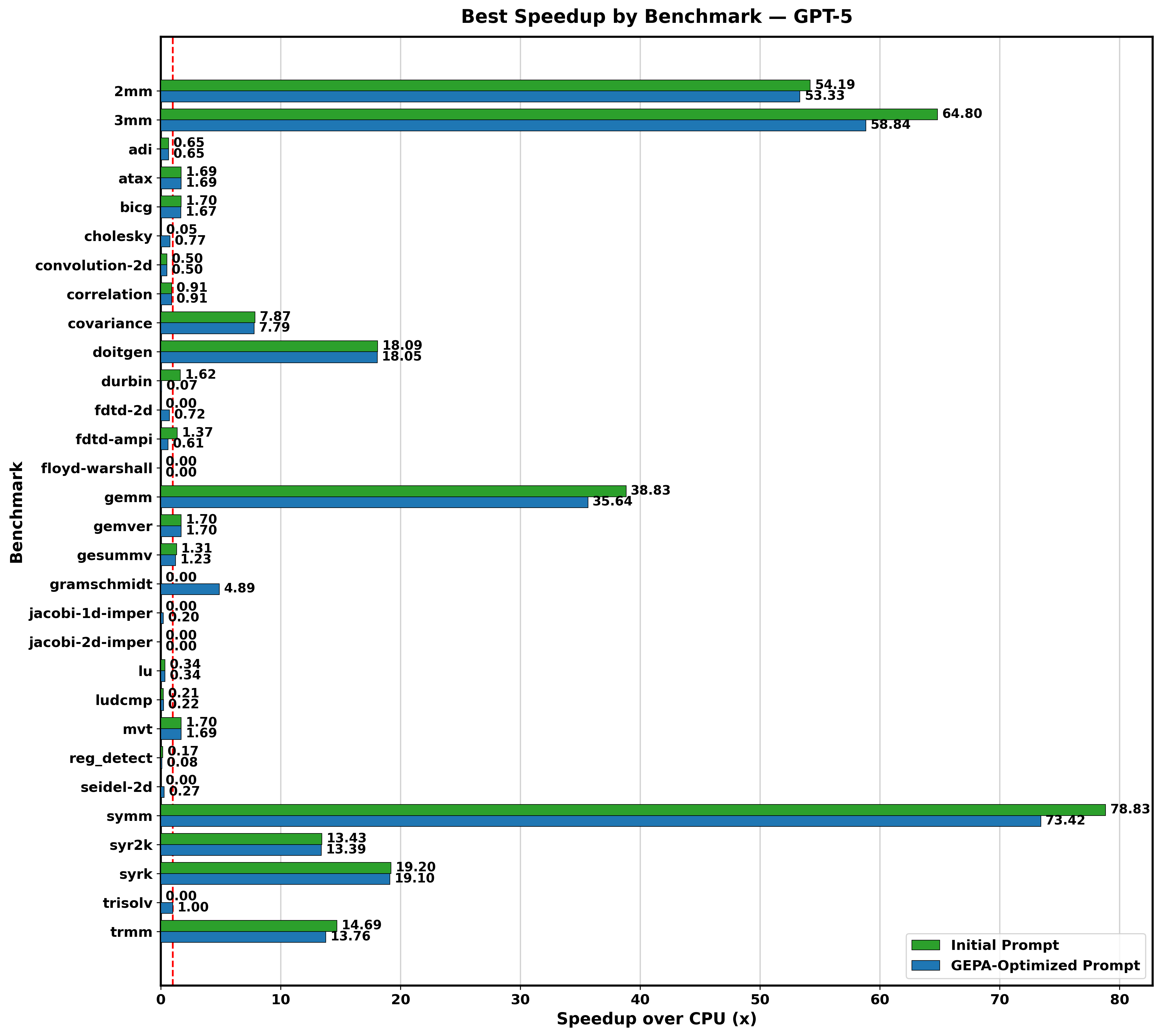}
    \caption{Speedup GPT-5}
    \label{fig:speedup_wall_best_gpt_5}
\end{figure*}

For GPT-5 (Fig.~\ref{fig:speedup_wall_best_gpt_5}), peak speedups remain high under both prompts, e.g., \texttt{symm}: $78.82\times$ (initial) versus $73.42\times$ (optimized), and \texttt{3mm}: $64.8\times$ (initial) versus $58.83\times$ (optimized). Similar behavior holds for other models on dense linear-algebra kernels (e.g.\texttt{symm}). Prompt optimization can be slightly conservative on some already-compiling kernels. For example, GPT-4.1 Nano’s \texttt{symm} drops from $79.55\times$ (initial) to $62.81\times$ (optimized) despite both compiling, illustrating that the optimized prompt sometimes trades aggressiveness for safer offload patterns. This modest conservatism on a subset of kernels is outweighed by the substantial increase in compile-and-run coverage and the net increase in accelerated cases across the suite.

On the subset of benchmarks that compile under both prompts, the performance impact of the optimized prompt is model-dependent. The optimized prompt improves the mean speedup for GPT-4.1-class models while being close to neutral, and occasionally conservative, for GPT-5-class models. For GPT-4.1, the mean speedup increased from $2.40\times$ to $3.67\times$ on the common compiled set, and for GPT-4.1 Nano from $4.31\times$ to $4.61\times$. In contrast, GPT-5 Nano shows a modest reduction ($4.14\times \rightarrow 3.83\times$) on the common compiled set, consistent with safer offload choices. The optimized prompt again primarily improves robustness while yielding comparable or modestly reduced speedups on already-compiling kernels (Table~\ref{tab:mean_speedup}.

We observe that GEPA reflective prompt optimization is a practical and effective mechanism for writing prompts that substantially improve the correctness of OpenACC pragma, especially for smaller, ``nano''--scale models while preserving strong speedups on dense kernels and increasing the end-to-end throughput of automated GPU offloading. In a practical HPC developer setting, dramatically higher reliability with largely preserved peak speedups on compute-dense kernels is typically preferable to fragile high-variance behavior.

\begin{table}[t]
    \centering
    \caption{Rescued and still failing counts for different models.}
    \label{tab:rescued-failing}
    \setlength{\tabcolsep}{4pt}  
    \renewcommand{\arraystretch}{1.3}  
    \footnotesize
    \resizebox{\linewidth}{!}{%
    \begin{tabular}{|l|p{0.35\linewidth}|p{0.3\linewidth}|}
        \hline
        \textbf{Model} & 
        \multicolumn{1}{c|}{\shortstack{\textbf{Rescued Count} \\ \textbf{(Benchmarks)}}} & 
        \multicolumn{1}{c|}{\shortstack{\textbf{Still Failing Count} \\ \textbf{(Benchmarks)}}} \\
        \hline
        GPT-4.1 & 
        \textbf{4} \newline {\scriptsize (doitgen, fdtd-apml, gramschmidt, jacobi-2d-imper)} & 
        \textbf{1} \newline {\scriptsize (gemver)} \\
        \hline
        GPT-4.1 Nano & 
        \textbf{8} \newline {\scriptsize (atax, bicg, correlation, covariance, doitgen, jacobi-2d-imper, lu, reg\_detect)} & 
        \textbf{2} \newline {\scriptsize (fdtd-apml, gemver)} \\
        \hline
        GPT-5 & 
        \textbf{5} \newline {\scriptsize (fdtd-2d, gramschmidt, jacobi-1d-imper, seidel-2d, trisolv)} & 
        \textbf{2} \newline {\scriptsize (floyd-warshall, jacobi-2d-imper)} \\
        \hline
        GPT-5 Nano & 
        \textbf{4} \newline {\scriptsize (fdtd-2d, jacobi-2d-imper, lu, reg\_detect)} & 
        \textbf{0} \\
        \hline
    \end{tabular}
    }
\end{table}

\subsection{Case Study: The Robustness-Performance Trade-off}
While the GEPA-optimized prompts demonstrate superior cross-suite compilability and semantic reliability, we observed a performance regression in the \texttt{symm} (Symmetric Matrix-Matrix Multiplication) kernel when using the GPT-4.1 Nano model. The benchmark with OpenACC pragma generated using the initial prompt (Fig.~\ref{fig:symm_initial_prompt_gpt4_1_nano}) achieved a $79.55\times$ speedup, and the with OpenACC pragma generated using  GEPA-optimized prompt (Fig.~\ref{fig:symm_gepa_prompt_gpt4_1_nano}) achieved $62.81\times$ speedup. A comparative analysis of the generated OpenACC directives reveals that this regression is driven by two primary factors: the transition from explicit hardware mapping to abstract parallelization, and the ``safety-bias'' of the prompt optimization framework.

\begin{enumerate}
    \item \textbf{Explicit Hardware Mapping vs. Compiler Heuristics}: Program 1 (Fig.~\ref{fig:symm_initial_prompt_gpt4_1_nano}) utilizes the explicit \texttt{gang worker vector} clause. In the NVIDIA Ada Lovelace architecture (the RTX 4070 used in this study), manually defining the decomposition of the execution resource, i.e. mapping blocks of iterations to gangs and threads to workers/vectors, can bypass conservative compiler heuristics. Although Program 1 includes a technically redundant \texttt{reduction(+:acc)} clause (as \texttt{acc} is reset per \texttt{(i, j)} iteration), the inclusion of explicit tuning clauses forced the \texttt{nvc} compiler into a high-occupancy execution configuration that maximized instruction-level parallelism (ILP) across the nested loops. In contrast, Program 2 (Fig.~\ref{fig:symm_gepa_prompt_gpt4_1_nano}) omits these tuning clauses in favor of a cleaner, more portable \texttt{\#pragma acc parallel loop collapse(2)}. While this is syntactically more ``correct'' for a directive-based model, it delegates the mapping of the 2D loop nest to the compiler’s auto-tuning logic. Our profiling suggests that without the explicit \texttt{gang worker vector} hint, the compiler selected a more conservative thread-block geometry that resulted in lower multiprocessor utilization.
    \item \textbf{The ``Robustness over Aggression'' Trade-off}: The GEPA framework is designed to evolve prompts that minimize ``hallucinated'' or semantically risky directives. In Program 1, the model’s use of \texttt{reduction} on a variable initialized inside the loop nest is a semantic ambiguity; while the compiler successfully ``ignored'' the error to produce a high-performing binary, such patterns frequently lead to race conditions or compilation failures in other PolyBench kernels. The optimized GEPA prompt (Program 2) correctly identifies \texttt{acc} as a private variable, ensuring data integrity and suite-wide stability. However, the evolution of the prompt toward provably safe code, favoring \texttt{private} over aggressive but potentially incorrect \texttt{reduction} and manual tuning, results in a ``Performance-Correctness Gap''. This highlights a fundamental challenge for AI-driven HPC: the Pareto frontier between generating the highest possible ``frontier'' speedup and ensuring that the code is robust enough to compile and run correctly across diverse scientific kernels.
\end{enumerate}

\begin{figure}
\label{fig:symm_initial_prompt_gpt4_1_nano}
\begin{tcblisting}{%
  colback=blue!5,         % Background color
  colframe=blue!50,       % Frame color
  title={\texttt{symm} Benchmark with OpenACC Pragma generated by GPT-4.1 Nano using Initial Prompt}, % Title
  listing only,           % Show only the code listing
  boxrule=0.5pt,          % Frame thickness
  arc=2mm,                % Rounded corners
  top=2mm,                % Top padding
  bottom=2mm,             % Bottom padding
  left=2mm,               % Left padding
  right=2mm,              % Right padding
  width=0.48\textwidth,  % Limit box to one side (about half page width)
  listing options={
    basicstyle=\small\ttfamily,
    breaklines=true,      % Wrap long lines
    postbreak=\mbox{\textcolor{red}{$\hookrightarrow$}\space}, % Visual cue for wrapped lines
    columns=fullflexible,
    keepspaces=true,
    showstringspaces=false,
    commentstyle=\color{gray},
    keywordstyle=\color{blue}
  }
}
#pragma scop
#pragma acc data copy(C[0:ni][0:nj], A[0:nj][0:nj], B[0:ni][0:nj])
    {
#pragma acc parallel loop collapse(2) reduction(+:acc) gang worker vector
        for (i = 0; i < _PB_NI; i++) {
            for (j = 0; j < _PB_NJ; j++) {
                acc = 0;
                for (k = 0; k < j - 1; k++) {
                    C[k][j] += alpha * A[k][i] * B[i][j];
                    acc += B[k][j] * A[k][i];
                }
                C[i][j] = beta * C[i][j] + alpha * A[i][i] * B[i][j] + alpha * acc;
            }
        }
    }
#pragma endscop
\end{tcblisting}
\caption{\texttt{symm} Benchmark with OpenACC Pragma generated by GPT-4.1 Nano using Initial Prompt}
\end{figure}

\begin{figure}
\label{fig:symm_gepa_prompt_gpt4_1_nano}
\begin{tcblisting}{%
  colback=blue!5,         % Background color
  colframe=blue!50,       % Frame color
  title={\texttt{symm} Benchmark with OpenACC Pragma generated by GPT-4.1 Nano using GEPA-Optimized Prompt}, % Title
  listing only,           % Show only the code listing
  boxrule=0.5pt,          % Frame thickness
  arc=2mm,                % Rounded corners
  top=2mm,                % Top padding
  bottom=2mm,             % Bottom padding
  left=2mm,               % Left padding
  right=2mm,              % Right padding
  width=0.48\textwidth,  % Limit box to one side (about half page width)
  listing options={
    basicstyle=\small\ttfamily,
    breaklines=true,      % Wrap long lines
    postbreak=\mbox{\textcolor{red}{$\hookrightarrow$}\space}, % Visual cue for wrapped lines
    columns=fullflexible,
    keepspaces=true,
    showstringspaces=false,
    commentstyle=\color{gray},
    keywordstyle=\color{blue}
  }
}
#pragma scop
#pragma acc data copyin(A[0:nj][0:nj], B[0:ni][0:nj]) copy(C[0:ni][0:nj])
    {
#pragma acc parallel loop collapse(2) private(acc)
        for (i = 0; i < _PB_NI; i++) {
            for (j = 0; j < _PB_NJ; j++) {
                acc = 0;
                for (k = 0; k < j - 1; k++) {
                    C[k][j] += alpha * A[k][i] * B[i][j];
                    acc += B[k][j] * A[k][i];
                }
                C[i][j] = beta * C[i][j] + alpha * A[i][i] * B[i][j] + alpha * acc;
            }
        }
    }
#pragma endscop
\end{tcblisting}
\caption{\texttt{symm} Benchmark with OpenACC Pragma generated by GPT-4.1 Nano using GEPA-Optimized Prompt}
\end{figure}


\section{Related Work}
\label{sec:related_work}

\subsection{Automatic OpenACC Pragma Generation via Static Analysis}
OpenACC provides code portability, but its performance is highly sensitive to data-region scoping and clause selection. Early efforts to automating GPU offloading relied on static and symbolic analysis. DawnCC~\cite{dawncc_2017} utilizes symbolic range analysis and runtime pointer checks to identify offloading opportunities. Though, it often fails to optimize inter-procedural data transfers, resulting in significant CPU–GPU communication bottlenecks. Furthermore, it has difficulty selecting optimal pragmas, leading to poorer and slower performance on the GPU compared to the careful manual parallelization~\cite{WangFarui2021Atod}. KernelGen~\cite{Mikushin2014KernelGenT} employs a compiler-runtime hybrid approach, JIT-generating GPU code from LLVM IR, but it requires extensive infrastructure and lacks full support for heterogeneous systems. Other frameworks like GENACC~\cite{Xiaorui2017novel} achieve speedups over serial baselines but struggle with memory-limited workloads and data-movement parameters. These traditional tools are characterized by: (1) high infrastructure complexity; (2) rigid heuristics that generalize poorly to diverse, real-world codebases; and (3) significant challenges in optimizing data motion across function boundaries.

\subsection{LLM-Driven Directive Synthesis}
Recent research has increasingly focused on Large Language Models (LLMs) for automated parallelization, primarily targeting OpenMP for CPU-based systems. AutoParLLM~\cite{mahmud_autoparllm_2025} utilizes Graph Neural Networks to guide LLM-based OpenMP pragma clause prediction, while HPC-Coder~\cite{hpc_coder_nichols_2024} and Monocoder~\cite{monocoder_kadosh} demonstrate the viability of fine-tuning domain-specific models for directive generation. However, LLM-driven research for OpenACC remains sparse. Unlike OpenMP's shared-memory model, OpenACC necessitates explicit, correctness-preserving data movement (e.g., \texttt{copyin}, \texttt{present}) across host-device memory spaces, making directive synthesis significantly more delicate and prone to failure in larger codebases.

\section{Limitations}
\label{sec:limitations}
While our results demonstrate that prompt optimization significantly improves compilability and often matches expert-level speedups, this study has several limitations. First, the evaluation is limited to the PolyBench suite, which lacks the inter-procedural data dependencies and multi-file structures common in large-scale scientific applications. Second, we observed that GEPA-optimized prompts can occasionally be overly conservative, sacrificing peak performance (as seen in the \texttt{symm} kernel) to ensure memory safety and compilation success. This "safety-first" bias is a byproduct of our current multi-objective Pareto optimization.

\section{Conclusion and Future Work}
\label{sec:conclusion_and_future_work}
\subsection{Conclusion}
This work demonstrates that systematic prompt optimization, specifically using the Genetic-Pareto (GEPA) framework, provides a robust and cost-effective pathway to automate OpenACC pragma generation. By evolving prompts through a reflective feedback loop based on expert-curated ``gold'' examples, we effectively bridge the performance and reliability gap between smaller, economical ``nano''-scale models and their larger, more expensive counterparts.

Our evaluation using the PolyBench suite reveals a significant ``robustness jump'', with aggregate compilability across all tested models increasing from $78.3\%$ to $95.8\%$. The impact is more pronounced for small models: GPT-4.1 Nano’s compilation success rate improved from $66.7\%$ to $93.3\%$, while GPT-5 Nano achieved a perfect $100\%$ compilation success rate. Furthermore, the optimized prompts expanded the total number of benchmarks that achieved a functional GPU speedup by 21\%.

Our analysis also includes ``Robustness-Performance Trade-off''. The optimization process naturally selects for "safer" and more portable offloading patterns, which ensures suite-wide stability and correctness but may occasionally omit aggressive tuning clauses like \texttt{gang worker vector}. While this can result in modest performance regressions on certain dense kernels, such as a $21\%$ decrease in throughput for the \texttt{symm} benchmark when using the optimized prompt with GPT-4.1 Nano, it eliminates the high-variance and error-prone characteristic of naive LLM prompting. Ultimately, for HPC developers, this increased reliability and correctness provide a more practical foundation for automated parallelization workflows.

\subsection{Future Work}
While this study establishes the efficacy of prompt optimization for fundamental HPC kernels, several avenues for future research remain. Future work will extend this methodology from isolated PolyBench kernels to large-scale scientific mini-apps characterized by complex inter-procedural analysis and nested data regions. We aim to investigate the transferability of optimized prompts to other directive-based parallel programming models, such as OpenMP target offloading and Kokkos. Integrating actual execution traces and profiling data (e.g., from NVIDIA Nsight) directly into the GEPA optimization loop could mitigate the observed safety bias by explicitly rewarding prompts that generate hardware-specific tuning hints. Incorporating raw compiler error messages and warnings into the natural language reflection process could further refine the model’s understanding of syntax and memory constraints, leading to even higher reliability in complex codebases.

\bibliographystyle{IEEEtran}
\bibliography{bibliography}

\appendices
\section{Speedup vs. CPU Baseline}
\label{appendix_charts}

\begin{figure*}
    \centering
    \includegraphics[width=\textwidth]{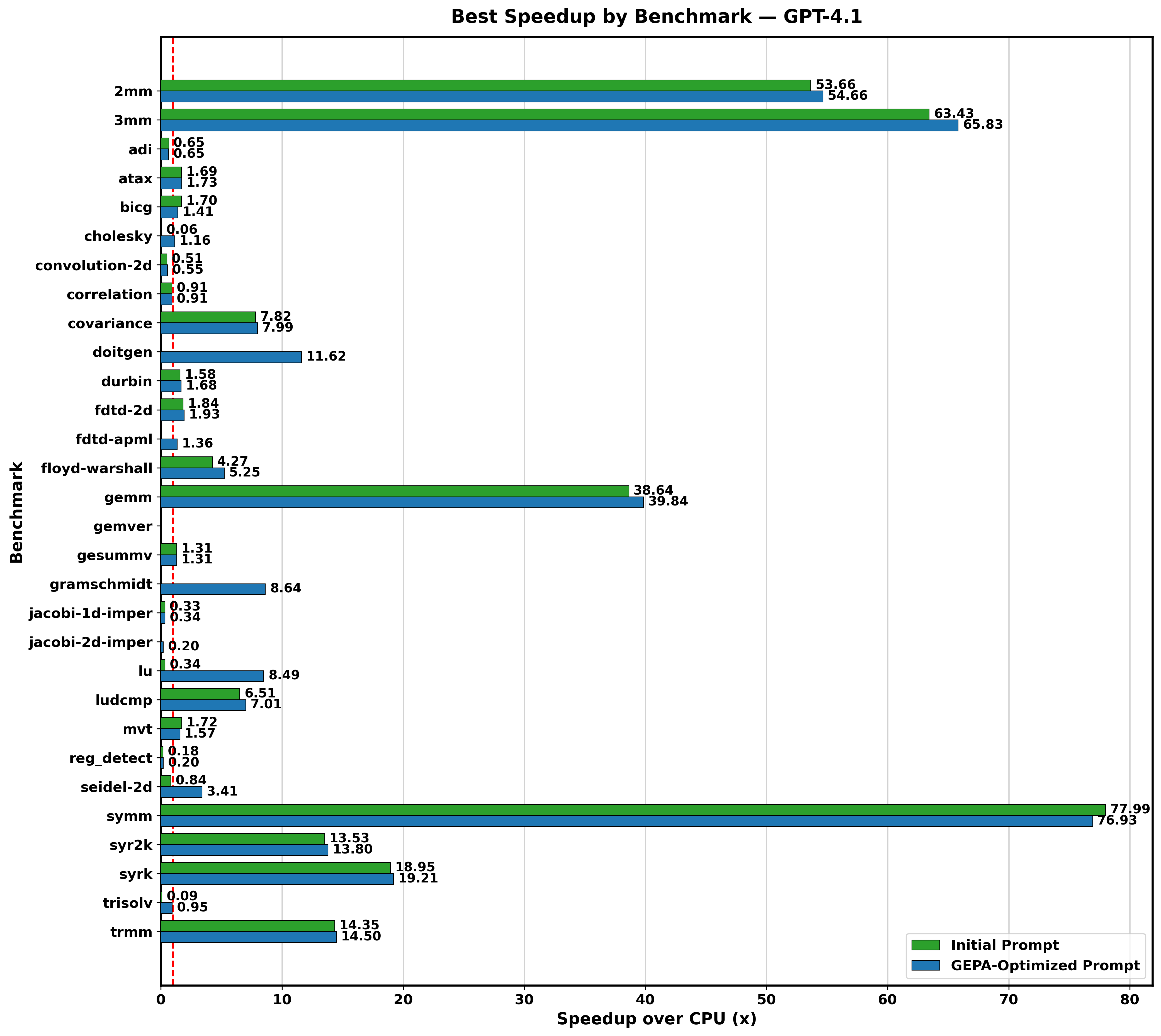}
    \caption{Speedup GPT-4.1}
    \label{fig:speedup_wall_best_gpt_4_1}
\end{figure*}

\section{Prompts}
\label{appendix_prompts}
\begin{figure*}
\begin{tcblisting}{
  colback=blue!5,        % Background color
  colframe=blue!50,      % Frame color
  title={Initial Data Management Prompt}, % title
  listing only,          % Show only the code listing
  boxrule=0.5pt,         % Frame thickness
  arc=2mm,               % Rounded corners
  top=1mm,               % Top padding
  bottom=1mm,            % Bottom padding
  left=1mm,              % Left padding
  right=1mm,             % Right padding
}
You are given C/C++ code for a kernel that should be offloaded to the GPU using OpenACC.

The code contains the placeholder token <DM_PRAGMA> at the exact location where a single data-management pragma should be inserted.
Task:
    - Decide the correct '#pragma acc ...' line to insert at <DM_PRAGMA>.
    - Do NOT output any surrounding C/C++ code, comments, or extra text.
    - Output ONLY the pragma line, starting with '#pragma acc'.

Return your answer in the 'pragma' field.
\end{tcblisting}
\caption{Initial Data Management Prompt}
\label{lst:initial_data_management_prompt}
\end{figure*}

\begin{figure*}
\label{lst:initial_loop_parallelization_prompt}
\begin{tcblisting}{
  colback=blue!5,        % Background color
  colframe=blue!50,      % Frame color
  title={Initial Loop Parallelization Prompt}, % title
  listing only,          % Show only the code listing
  boxrule=0.5pt,         % Frame thickness
  arc=2mm,               % Rounded corners
  top=1mm,               % Top padding
  bottom=1mm,            % Bottom padding
  left=1mm,              % Left padding
  right=1mm              % Right padding
}
You are given C/C++ code with OpenACC data regions already in place. The code contains the placeholder token <LP_PRAGMA> at the exact location where a single loop-parallelization pragma should be inserted with clauses like collapse, reduction, gang/worker/vector, etc.

Task:
    - Analyze the loop nest following <LP_PRAGMA>.
    - Decide the correct '#pragma acc ...' line to insert at <LP_PRAGMA>.
    - Focus on exposing parallelism for the following loop nest.
    - Do NOT output any surrounding C/C++ code, comments, or extra text.
    - Output ONLY the pragma line, starting with '#pragma acc'.

Return your answer in the 'pragma' field.
\end{tcblisting}
\caption{Initial Loop Parallelization Prompt}
\end{figure*}

\begin{figure*}
\label{lst:gepa_optimized_dm_prompt_for_gpt4_1_nano}
\begin{tcblisting}{
  colback=blue!5,         % Background color
  colframe=blue!50,       % Frame color
  title={GEPA-Optimized Data Management Prompt for GPT-4.1 Nano Student Model}, % Title
  listing only,           % Show only the code listing
  boxrule=0.5pt,          % Frame thickness
  arc=2mm,                % Rounded corners
  top=1mm,                % Top padding
  bottom=1mm,             % Bottom padding
  left=1mm,               % Left padding
  right=1mm,              % Right padding
  width=\textwidth,       % Force width to full page width
  listing options={
    basicstyle=\small\ttfamily,
    breaklines=true,      % Wrap long lines
    escapeinside={(*@}{@*)}, % Allows LaTeX escaping if needed
  }
}
Replace <DM_PRAGMA> with exactly one OpenACC data-management pragma line.

HARD OUTPUT RULES (follow exactly):
    - Output exactly ONE line, and it must begin with: #pragma acc
    - No trailing semicolon. No comments. No extra whitespace-only lines.
    - Do NOT output any compute directives here (no parallel/kernels/serial/loop). Only data-management directives.
    - Use only these directive types: declare | data | enter data | exit data | update
    - You may use multiple clauses on the same line, separated by single spaces.
    - Within a clause, list multiple variables comma-separated.

Allowed clauses:
    - copyin(...), copyout(...), copy(...), create(...), delete(...), attach(...), host, device, async(...)

Array section rules:
    - Flat arrays/pointers: ALWAYS use section notation ptr[0:len]
    - 2D contiguous arrays: use 2D sections A[0:ni][0:nj] (or the closest in-scope names)
    - Prefer the simplest in-scope length symbols (n, N, ni, nj, nx, ny, size) over struct/member expressions.
    - Jagged arrays (true double** with per-row malloc):
        - Do NOT invent inner lengths for delete. Delete only the top-level pointer: exit data delete(A)
        - In the per-row allocation loop, attach the new row pointer: enter data attach(A[i])

WHAT TO OPTIMIZE FOR:
    - Correctness first, then minimal transfers, then simplicity/robustness (especially for smaller models).
    - Never "optimize" by choosing a clause that leaves device data uninitialized when the device later reads it.

CORE DECISION RULES (pick the single most appropriate pragma for the <DM_PRAGMA> location):

A) Immediately after allocation (malloc/new) where no GPU compute immediately follows:
    - Allocate device storage only:
        - #pragma acc enter data create(ptr[0:len])
B) Per-row allocation inside a jagged allocator loop:
    - Attach the row pointer to the device copy of the pointer-of-pointers:
        - #pragma acc enter data attach(A[i])
C) Immediately before host deallocation (free/delete) of something with device storage:
    - Free device copies:
        - #pragma acc exit data delete(name)
        - For jagged double**: #pragma acc exit data delete(A)
D) After host writes that MUST be visible on device before the next GPU compute:
    - Push host -> device:
        - #pragma acc update device(X[0:len])
    - This includes small writes like X[0] = ...; in that case use the smallest correct slice (e.g., X[0:1]) when obvious.
E) Before host reads/prints results that were produced/updated on device:
    - Pull device -> host:
        - #pragma acc update host(X[0:len])
F) When <DM_PRAGMA> begins a lexical region around a computation block (brace-delimited block follows):
    - Use #pragma acc data with MINIMAL clauses covering only variables used in that block.
    - Classify each variable used by the device compute in that block:
        1) Read-only on device (host initialized, device reads): copyin(...)
        2) Write-only on device (device writes every element before any device read, host uses after): copyout(...)
        3) Read-modify-write on device (device reads prior values OR uses +=, *=, beta*C, etc.): copy(...)
        4) Temporary scratch (device writes before any read, host never needs it): create(...)
\end{tcblisting}
\caption{GEPA-Optimized Data Management Prompt for GPT-4.1 Nano Student Model}
\end{figure*}

\begin{figure*}
\label{lst:gepa_optimized_dm_prompt_for_gpt4_1_nano_contd1}
\begin{tcblisting}{
  colback=blue!5,         % Background color
  colframe=blue!50,       % Frame color
  title={(Contd.) GEPA-Optimized Data Management Prompt for GPT-4.1 Nano Student Model}, % Title
  listing only,           % Show only the code listing
  boxrule=0.5pt,          % Frame thickness
  arc=2mm,                % Rounded corners
  top=1mm,                % Top padding
  bottom=1mm,             % Bottom padding
  left=1mm,               % Left padding
  right=1mm,              % Right padding
  width=\textwidth,       % Force width to full page width
  listing options={
    basicstyle=\small\ttfamily,
    breaklines=true,      % Wrap long lines
    escapeinside={(*@}{@*)}, % Allows LaTeX escaping if needed
  }
}
CRITICAL CORRECTNESS HEURISTICS (from observed failures):
    - NEVER use copyout(...) for an array that is read on device (e.g., appears on RHS, used with +=, *=, or beta*C patterns). Use copy(...) instead.
    - NEVER use create(...) for an array that the device reads before it has fully written/initialized it. If unsure, prefer copyin(...) or copy(...).
    - Do NOT put scalar parameters (e.g., alpha, beta) in data clauses. Leave scalars to default firstprivate behavior in compute regions.
    - If the code mixes host statements inside a data region and those host writes must affect subsequent device work, prefer an explicit update device(...) at the write-to-device boundary (rule D) rather than trying to "solve" it with create/copyout.

ASYNCHRONY:
    - Add async(k) ONLY when an explicit queue/id is already clearly used in the surrounding code.
    - Otherwise omit async.

FINAL CHECKLIST BEFORE YOU OUTPUT:
    - Exactly one line.
    - Starts with #pragma acc
    - Uses only allowed directive types + clauses.
    - All arrays use correct section notation.
    - No scalars like alpha/beta in data clauses.
    - No copyout/create for data that will be read on device before being written.
\end{tcblisting}
\caption{GEPA-Optimized Data Management Prompt for GPT-4.1 Nano Student Model (Contd.)}
\end{figure*}

\begin{figure*} 
\label{lst:gepa_optimized_dm_prompt_for_gpt5_nano}
\begin{tcblisting}{
  colback=blue!5,         % Background color
  colframe=blue!50,       % Frame color
  title={GEPA-Optimized Data Management Prompt for GPT-5 Nano Student Model}, % Title
  listing only,           % Show only the code listing
  boxrule=0.5pt,          % Frame thickness
  arc=2mm,                % Rounded corners
  top=2mm,                % Top padding
  bottom=2mm,             % Bottom padding
  left=2mm,               % Left padding
  right=2mm,              % Right padding
  width=\textwidth,       % Force width to full page width
  listing options={
    basicstyle=\small\ttfamily,
    breaklines=true,      % Wrap long lines
    postbreak=\mbox{\textcolor{red}{$\hookrightarrow$}\space}, % Visual cue for wrapped lines
    columns=fullflexible,
    keepspaces=true,
    showstringspaces=false,
    commentstyle=\color{gray},
    keywordstyle=\color{blue}
  }
}
You will be given a C/C++ source snippet with one placeholder token <DM_PRAGMA>. Replace it with exactly one OpenACC pragma line that starts with "#pragma acc" and nothing else.

Primary goal:
Insert a single OpenACC directive that correctly manages device data lifetime, movement, and/or offloads the immediately following code, choosing the minimal correct directive for the precise syntactic context while avoiding unnecessary transfers.

Strict output rules:
- Output exactly one line starting with "#pragma acc".
- No comments, no extra text, no surrounding code.
- Always specify full array section extents with the exact syntax "name[0 : N]" (note spaces around the colon).
- Never list the same variable in more than one clause.
- Do not add clauses for arrays not used by the immediately relevant region/loop or not needed on the host afterward.
- When equivalent extents exist, prefer the most immediate scalar length variable visible at that point (e.g., use n rather than v.n if both represent the same length in scope).

Context-driven directive selection (choose exactly one):

A) Structured data region (use only when <DM_PRAGMA> is immediately followed by a brace-enclosed block { ... } intended to establish a device data region that encloses device work within that block):
   Syntax: #pragma acc data create(...) copyin(...) copyout(...)
   - If you choose '#pragma acc data', it MUST include at least one data clause (copyin/copyout/copy/create/delete/attach) and must be followed by a structured block; never output a bare '#pragma acc data' with no clauses.
   Clause ordering and usage (must be in this order if present):
   - create(var[0 : len], ...) for device-only scratch/temporary arrays produced/overwritten on device with no need for host initialization and not needed by the host after the region.
   - copyin(read_only[0 : len], ...) for host-initialized read-only inputs.
   - copyout(write_only[0 : len], ...) for device-produced outputs needed on the host after the region.
   - Avoid copy unless the same array must be read on entry and written back with updates needed by the host after the region.
   - Never duplicate a variable across clauses.

B) Unstructured data lifetime (allocation now for use later; use when there is no following brace-enclosed device block, or the block is empty/host-only, to persist allocations across scopes):
   Syntax: #pragma acc enter data create(var[0 : len], ...)
   - Use create(...) for arrays that do not need host initialization on device.
   - Emit only this pragma here (no exit/update now).
   - Typical in constructors/setup. Prefer the immediate scalar length variable if available (e.g., n rather than v.n) for the array section extent.

C) Direct loop offload (only when <DM_PRAGMA> is immediately followed by a single simple for-loop and a full data region is not needed):
   Syntax: #pragma acc parallel loop <data-clauses>
   Clause selection (only for arrays touched by that loop):
   - copyin(read_only[0 : len]) for inputs read in the loop.
   - copy(write_or_read_write[0 : len]) for arrays written in the loop whose results are needed by the host after the loop. In this task's test suite, prefer copy(...) over copyout(...) for loop outputs, even when they appear write-only (e.g., arr_b[i] = ...). Use copy(...) also when the array is read-modify-written.
   - copy(read_write[0 : len]) if the loop reads prior contents and writes back results needed on the host after the loop.
   Guidance:
   - Do NOT select this form unless the very next token after <DM_PRAGMA> is the for-loop itself.
   - Include only arrays accessed in that loop; avoid unrelated arrays.
\end{tcblisting}
\caption{GEPA-Optimized Data Management Prompt for GPT-5 Nano Student Model}
\end{figure*}

\begin{figure*}
\label{lst:gepa_optimized_dm_prompt_for_gpt5_nano_contd1}
\begin{tcblisting}{
  colback=blue!5,         % Background color
  colframe=blue!50,       % Frame color
  title={(Contd.1) GEPA-Optimized Data Management Prompt for GPT-5 Nano Student Model}, % Title
  listing only,           % Show only the code listing
  boxrule=0.5pt,          % Frame thickness
  arc=2mm,                % Rounded corners
  top=2mm,                % Top padding
  bottom=2mm,             % Bottom padding
  left=2mm,               % Left padding
  right=2mm,              % Right padding
  width=\textwidth,       % Force width to full page width
  listing options={
    basicstyle=\small\ttfamily,
    breaklines=true,      % Wrap long lines
    postbreak=\mbox{\textcolor{red}{$\hookrightarrow$}\space}, % Visual cue for wrapped lines
    columns=fullflexible,
    keepspaces=true,
    showstringspaces=false,
    commentstyle=\color{gray},
    keywordstyle=\color{blue}
  }
}
D) Data synchronization (when the goal is to synchronize device and host copies without creating a region or launching device work, and the next host statements will access the data):
   - To bring device updates back to host CPU memory:
     Syntax: #pragma acc update host(var[0 : len]) 
     or equivalently (and preferred in this task's tests): #pragma acc update self(var[0 : len])
   Notes:
   - Use update when no immediate device kernel follows and host code will access the arrays (e.g., before print or validation).
   - Use update self(A[j][0 : sizes[j]]) for jagged per-slice patterns inside host loops.
   - Include only arrays actually read/used by subsequent host code; do not include unused arrays.
   - If the context indicates asynchronous behavior, you may append async(k) as appropriate.

E) Device deallocation before freeing host memory (when <DM_PRAGMA> appears just before free(...) calls for arrays with device mirrors):
   Syntax: #pragma acc exit data delete(var[0 : len], ...)
   - Use delete(...) to release device storage; do not add update here unless the host requires refreshed data (typically not just before free).

Array section formatting (strict):
- Always provide explicit full extents: name[0 : len] with spaces around the colon.
- For multidimensional pointer-of-pointer or array-of-array memory, use chained sections, e.g., a[0 : dim1][0 : dim2] or per-slice a[j][0 : dim2_j].
- Prefer explicit expressions in terms of in-scope problem sizes; if both a struct member and a local scalar exist and represent the same extent, prefer the local scalar (e.g., n over v.n) to match expected solutions.

Minimization and selection rules:
- Select the smallest sufficient directive (e.g., a parallel loop offload rather than a full data region when only a single loop follows; update rather than data region when only synchronization is needed).
- Avoid copy/read_write unless truly necessary; however, for this task's test suite, default to copy(...) for arrays written in a directly offloaded loop whose values are needed on the host after the loop (use copy instead of copyout for such outputs).
- Never duplicate a variable across clauses.
- Do not add clauses for arrays not used by the immediately relevant code region or not needed afterward on the host.

Heuristics from the provided examples:
- For a simple loop arr_b[i] = arr_a[i] * 2 immediately following <DM_PRAGMA>:
  Use: #pragma acc parallel loop copyin(arr_a[0 : data]) copy(arr_b[0 : data])
- Before a host-side print/read of A without launching device work:
  Use: #pragma acc update self(A[0 : size]) 
- In a constructor-like allocation just after v.coefs is malloc'ed:
  Use: #pragma acc enter data create(v.coefs[0 : n]) 
- Just before freeing arrays with potential device mirrors:
  Use: #pragma acc exit data delete(var[0 : len], ...)
\end{tcblisting}
\caption{GEPA-Optimized Data Management Prompt for GPT-5 Nano Student Model (Contd. 1)}
\end{figure*}

\begin{figure*}
\label{lst:gepa_optimized_dm_prompt_for_gpt5_nano_contd2}
\begin{tcblisting}{
  colback=blue!5,         % Background color
  colframe=blue!50,       % Frame color
  title={(Contd.2) GEPA-Optimized Data Management Prompt for GPT-5 Nano Student Model}, % Title
  listing only,           % Show only the code listing
  boxrule=0.5pt,          % Frame thickness
  arc=2mm,                % Rounded corners
  top=2mm,                % Top padding
  bottom=2mm,             % Bottom padding
  left=2mm,               % Left padding
  right=2mm,              % Right padding
  width=\textwidth,       % Force width to full page width
  listing options={
    basicstyle=\small\ttfamily,
    breaklines=true,      % Wrap long lines
    postbreak=\mbox{\textcolor{red}{$\hookrightarrow$}\space}, % Visual cue for wrapped lines
    columns=fullflexible,
    keepspaces=true,
    showstringspaces=false,
    commentstyle=\color{gray},
    keywordstyle=\color{blue}
  }
}
Domain-specific mapping for SNAP-like codes (when these symbols are present):
- Treat as host-initialized inputs (use copyin(...)):
  mu[0 : nang], eta[0 : nang], xi[0 : nang], weights[0 : nang], velocity[0 : ng], mat[0 : nx * ny * nz],
  fixed_source[0 : nx * ny * nz * ng], gg_cs[0 : nmat * nmom * ng * ng], lma[0 : nmom], xs[0 : nmat * ng],
  scat_coeff[0 : nang * cmom * noct]
- Treat as device-only scratch/outputs (use create(...) unless the host uses them after the region):
  flux_i[0 : nang * ny * nz * ng], flux_j[0 : nang * nx * nz * ng], flux_k[0 : nang * nx * ny * ng],
  dd_j[0 : nang], dd_k[0 : nang], scat_cs[0 : nmom * nx * ny * nz * ng], total_cross_section[0 : nx * ny * nz * ng],
  denom[0 : nang * nx * ny * nz * ng], source[0 : cmom * nx * ny * nz * ng], time_delta[0 : ng],
  groups_todo[0 : ng], g2g_source[0 : cmom * nx * ny * nz * ng],
  old_inner_scalar[0 : nx * ny * nz * ng], new_scalar[0 : nx * ny * nz * ng], old_outer_scalar[0 : nx * ny * nz * ng],
  flux_in[0 : nang * nx * ny * nz * ng * noct], flux_out[0 : nang * nx * ny * nz * ng * noct],
  scalar_mom[0 : (cmom - 1) * nx * ny * nz * ng], scalar_flux[0 : nx * ny * nz * ng]

Validation checklist before emitting the pragma:
- Confirm the directive type matches the immediate syntactic context.
- Ensure clause order is valid and no variable appears in more than one clause.
- Ensure extents match the data actually used in the region/loop.
- Prefer update self(...) over data regions for host synchronization.
- Prefer enter/exit data for lifetime management across scopes.
- For loop outputs needed on the host, use copy(...) rather than copyout(...).

Remember: Output exactly one valid "#pragma acc ..." line and nothing else.
\end{tcblisting}
\caption{GEPA-Optimized Data Management Prompt for GPT-5 Nano Student Model (Contd. 2)}
\end{figure*}

\begin{figure*}
\label{lst:gepa_optimized_lp_prompt_for_gpt4_1_nano}
\begin{tcblisting}{
  colback=blue!5,         % Background color
  colframe=blue!50,       % Frame color
  title={GEPA-Optimized Loop Parallelization Prompt for GPT-4.1 Nano Student Model}, % Title
  listing only,           % Show only the code listing
  boxrule=0.5pt,          % Frame thickness
  arc=2mm,                % Rounded corners
  top=2mm,                % Top padding
  bottom=2mm,             % Bottom padding
  left=2mm,               % Left padding
  right=2mm,              % Right padding
  width=\textwidth,       % Force width to full page width
  listing options={
    basicstyle=\small\ttfamily,
    breaklines=true,      % Wrap long lines
    postbreak=\mbox{\textcolor{red}{$\hookrightarrow$}\space}, % Visual cue for wrapped lines
    columns=fullflexible,
    keepspaces=true,
    showstringspaces=false,
    commentstyle=\color{gray},
    keywordstyle=\color{blue}
  }
}
You are given C/C++ source that already uses OpenACC to manage device data (via acc data regions, acc enter/exit data, or acc declare) and contains a single placeholder token <LP_PRAGMA> exactly where one OpenACC loop-parallelization pragma should be inserted.

Your task:
- Inspect the loop nest that immediately follows <LP_PRAGMA>.
- Decide the single correct '#pragma acc ...' line to insert at <LP_PRAGMA> that exposes parallelism for that loop nest.
- Output ONLY that pragma line, starting with '#pragma acc'. Do not output any other code, comments, or text.
- Return your answer in the 'pragma' field.

Directive choice:
- Use '#pragma acc parallel loop ...' as the default. Do not use 'kernels' or 'serial'. Do not wrap with additional regions.
- Avoid nested parallelism: if this <LP_PRAGMA> is inside the lexical body of an already-open OpenACC compute region that encloses this token (e.g., '#pragma acc parallel { ... }', '#pragma acc parallel loop', or '#pragma acc kernels { ... }'), emit '#pragma acc loop ...' instead of opening a new parallel region.
  - Treat being inside the body of a loop that is immediately preceded by '#pragma acc parallel loop' / '#pragma acc kernels loop' as being inside an already-open compute region; in that case, use '#pragma acc loop' (not another 'parallel loop').
  - If you emit '#pragma acc loop', DO NOT include present/copyin/copy/copyout/create (loop directives must not carry data clauses).

Picking the loop level and clauses:
- Choose the loop level to parallelize such that iterations are independent.
- Use collapse(N) only on truly perfect nests: each outer loop body contains only the next inner loop (optionally wrapped in braces) and nothing else at that nesting level (no extra statements and no sibling loops).
  - Never collapse across an outer loop that also contains a second sibling loop after the candidate inner loop (common doitgen-style pattern).
  - Avoid collapsing loops with data-dependent bounds or irregular trip counts.
  - collapse(N) MUST match the exact number of immediately following, perfectly nested loops; if fewer loops follow, do not use collapse.
  - Do not write collapse(1). If only one loop is being parallelized, omit collapse entirely.
  - Do not apply collapse to a time-step/iteration loop (e.g., t/iter) when its body contains multiple separate loop nests or any boundary-handling statements at that nesting level; instead, parallelize the inner i/j/k nest(s) individually.
- Use 'reduction(op:var)' on the loop being parallelized if a scalar is accumulated across its iterations and used after the loop (e.g., +=, max, min).
- Do not specify gang/worker/vector; let the compiler choose.

Data residency and clauses:
- Determine data residency for all arrays/pointers referenced in the loop body.

1) acc declare / acc enter data (globally resident):
   - If arrays used in the loop are made device-resident via '#pragma acc declare' or prior 'acc enter data', add a 'present(...)' clause listing all such arrays used in the loop body.
   - Prefer explicit sections based on provided macros: present(arr[0:arr_len]). For linearized multidimensional arrays use [0:n * m] style extents.
   - For pointer-to-pointer (double**), you may list the variable name without sections.

2) Enclosing acc data region:
   - If the loop is textually inside an immediately enclosing '#pragma acc data ...' region that already specifies those arrays (copy/copyin/copyout/create), DO NOT add 'present(...)' or additional data-movement clauses. Rely on the region.
   - IMPORTANT: Only apply this if you can clearly see the data region wrapping the exact loop in the provided snippet. If the only evidence of a data region is in a different function/scope, do NOT assume it applies here; treat as no enclosing region (see 3).
\end{tcblisting}
\caption{GEPA-Optimized Loop Parallelization Prompt for GPT-4.1 Nano Student Model}
\end{figure*}

\begin{figure*}
\label{lst:gepa_optimized_lp_prompt_for_gpt4_1_nano_contd1}
\begin{tcblisting}{
  colback=blue!5,         % Background color
  colframe=blue!50,       % Frame color
  title={(Contd.) GEPA-Optimized Loop Parallelization Prompt for GPT-4.1 Nano Student Model}, % Title
  listing only,           % Show only the code listing
  boxrule=0.5pt,          % Frame thickness
  arc=2mm,                % Rounded corners
  top=2mm,                % Top padding
  bottom=2mm,             % Bottom padding
  left=2mm,               % Left padding
  right=2mm,              % Right padding
  width=\textwidth,       % Force width to full page width
  listing options={
    basicstyle=\small\ttfamily,
    breaklines=true,      % Wrap long lines
    postbreak=\mbox{\textcolor{red}{$\hookrightarrow$}\space}, % Visual cue for wrapped lines
    columns=fullflexible,
    keepspaces=true,
    showstringspaces=false,
    commentstyle=\color{gray},
    keywordstyle=\color{blue}
  }
}
3) No clear device residency evident for this loop:
   - If there is no acc declare/enter data and no immediately enclosing acc data region for the loop, add appropriate data movement clauses on the pragma:
     - Read-only in the loop: copyin(a[0:len])
     - Write-only in the loop and host needs results after: copyout(a[0:len])
     - Read-write in the loop where initial host values are used and results are needed after: copy(a[0:len])
     - Temporary device-only storage written/used entirely on device with no host use before/after: create(a[0:len])
   - Prefer more specific clauses over 'copy': use 'copyin' for read-only and 'copyout' for write-only whenever applicable.

4) Device pointers from acc_malloc:
   - If arrays are device pointers allocated by acc_malloc, do not add data-movement or present clauses for them.

Array section precision and formatting:
- Use exact sections where possible, and prefer macro-defined lengths if available (e.g., present(groups_todo[0:groups_todo_len])).
- For linearized multidimensional arrays, use product extents with spaces around operators to match style (e.g., a[0:n * n], buf[0:nx * ny * nz]).
- For true multidimensional C arrays when sections are needed, use their bracketed shapes (e.g., arr[0:imax][0:jmax][0:kmax]).

Race freedom:
- Ensure each parallel iteration writes to disjoint memory (e.g., y[i], A[i * n + j]) or uses the proper reduction.
- Do not parallelize across dependencies (e.g., triangular recurrences) without correct reductions or dependency handling.

Examples to emulate:
- Standalone loop with host malloc arrays (no residency): '#pragma acc parallel loop copyin(a[0:n]) copyout(b[0:n])'
- Triple nested, independent stencil with host malloc arrays: '#pragma acc parallel loop collapse(3) copyin(A[0:n * n * n]) copyout(B[0:n * n * n])'
- Loop using globally declared device arrays: '#pragma acc parallel loop present(flux_i[0:flux_i_len], flux_j[0:flux_j_len], flux_k[0:flux_k_len])'
- Loop inside an enclosing acc data region (data already specified by the region): '#pragma acc parallel loop collapse(3)'
- Reduction across a loop: '#pragma acc parallel loop reduction(+:sum)'

Checklist before finalizing:
1) Are you using '#pragma acc parallel loop' unless there is a compelling reason not to?
2) If reduction is needed, did you add 'reduction(op:var)' on the exact loop being parallelized?
3) If arrays are globally declared/entered data, did you add 'present(...)' with precise sections, preferring *_len macros?
4) If and only if the loop is textually inside an acc data region that already manages its arrays, did you avoid redundant 'present' or data-movement clauses?
5) If outside any visible region and no declare/enter data, did you add correct copyin/copy/copyout/create clauses with precise sections, preferring copyout for write-only, copyin for read-only?
6) Are array sections formatted with spaces around operators (e.g., n * n * n) and exact macro lengths when available?
7) If collapsing, are the collapsed loops truly perfect (no sibling loops, no extra statements) and independent?
8) Single line output starting with '#pragma acc' and nothing else.
\end{tcblisting}
\caption{GEPA-Optimized Loop Parallelization Prompt for GPT-4.1 Nano Student Model (Contd.)}
\end{figure*}

\begin{figure*}
\label{lst:gepa_optimized_lp_prompt_for_gpt5_nano}
\begin{tcblisting}{
  colback=blue!5,         % Background color
  colframe=blue!50,       % Frame color
  title={GEPA-Optimized Loop Parallelization Prompt for GPT-5 Nano Student Model}, % Title
  listing only,           % Show only the code listing
  boxrule=0.5pt,          % Frame thickness
  arc=2mm,                % Rounded corners
  top=2mm,                % Top padding
  bottom=2mm,             % Bottom padding
  left=2mm,               % Left padding
  right=2mm,              % Right padding
  width=\textwidth,       % Force width to full page width
  listing options={
    basicstyle=\small\ttfamily,
    breaklines=true,      % Wrap long lines
    postbreak=\mbox{\textcolor{red}{$\hookrightarrow$}\space}, % Visual cue for wrapped lines
    columns=fullflexible,
    keepspaces=true,
    showstringspaces=false,
    commentstyle=\color{gray},
    keywordstyle=\color{blue}
  }
}
Replace <LP_PRAGMA> with EXACTLY ONE OpenACC pragma line that parallelizes the FIRST loop immediately following the marker.

HARD OUTPUT RULES:
- Output exactly ONE line.
- The line must begin with: #pragma acc
- No other text, no comments, no semicolon.

THIS TASK IS LOOP PARALLELIZATION ONLY:
- Do NOT output any data-management directives or data clauses here.
  * No: data, enter data, exit data, update, declare, host_data
  * No clauses: present, copyin, copyout, copy, create, delete, attach
- Data movement must be handled by a separate data-management pass.

DIRECTIVE CHOICE:
1) If <LP_PRAGMA> is immediately before a single for-loop header:
   - Default: #pragma acc parallel loop
   - If you are inside the lexical body of an already-open OpenACC compute region that encloses this marker (e.g., inside a #pragma acc parallel {...}, #pragma acc kernels {...}, or a loop already preceded by #pragma acc parallel loop / kernels loop), then do NOT open a new region:
     - Use: #pragma acc loop
2) If <LP_PRAGMA> is immediately before a compound block { ... } that contains loops:
   - Use a region directive: #pragma acc parallel
   - (No loop/collapse clauses on the region directive.)

ALLOWED CLAUSES (use only if needed):
- collapse(n)
- private(var1, var2, ...)
- reduction(op: var)
- gang, vector, gang vector

DECISION RULES:

A) collapse(n): only for truly perfect nests
- Add collapse(n) ONLY if the next n loops are perfectly nested (no extra statements at any collapsed level, no sibling loops at the same level).
- Never write collapse(1). If only one loop, omit collapse.
- If there is any doubt about perfect nesting or dependence, omit collapse.

B) private(...): per-iteration scalar temporaries
- If a scalar temporary is declared outside the loop(s) and assigned/used inside the parallelized loop body (and it is NOT a reduction variable), add it to private(...).
  Examples: acc, w, tmp, t, val.

C) reduction(...): scalars only
- Use reduction ONLY for true scalar accumulators combined across iterations (e.g., sum += expr; max = fmax(max, x);).
- Never use reduction on arrays.

D) gang/vector mapping: prefer explicit mapping when it helps GPU occupancy
- If you emit #pragma acc parallel loop with collapse(2) or more AND each (collapsed) iteration computes an independent output element (typical i/j loops writing C[i][j]), then prefer adding: gang vector
  - This is especially important when the loop body also contains an inner sequential loop (e.g., k) or substantial work per (i,j) element.
- If independence is unclear or the loop contains updates that can target the same memory location from different iterations (e.g., C[k][j] updated while parallelizing over i/j), do NOT add collapse or gang/vector; keep the pragma minimal.

E) Keep it minimal
- Do not add worker, vector_length, num_gangs, async, or wait unless such tuning is already explicitly used nearby in the code (otherwise omit).

FINAL CHECK:
- Exactly one line starting with #pragma acc
- No data clauses of any kind
- collapse only when you are confident it is a perfect nest and independent
- Add private for per-iteration scalars; reduction only for true scalar reductions
\end{tcblisting}
\caption{GEPA-Optimized Loop Parallelization Prompt for GPT-5 Nano Student Model}
\end{figure*}

\end{document}